\documentclass[runningheads]{llncs}
\usepackage{graphicx}
\usepackage{comment}
\usepackage{amsmath,amssymb} 
\usepackage{color}
\usepackage{multirow}
\usepackage{multicol}
\usepackage{caption}
\usepackage{bm}
\usepackage{mathrsfs}
\usepackage{booktabs}

\newlength\savewidth\newcommand\shline{\noalign{\global\savewidth\arrayrulewidth
  \global\arrayrulewidth 1pt}\hline\noalign{\global\arrayrulewidth\savewidth}}

\begin{document}
\pagestyle{headings}
\mainmatter
\def\ECCVSubNumber{1354}  

\def\ie{\emph{i.e.}}
\def\eg{\emph{e.g.}}
\def\etc{\emph{etc}}

\title{GIQA: Generated Image Quality Assessment} 

%
\author{
Shuyang Gu$^{1}$, Jianmin Bao$^{2}$\thanks{Corresponding author.}, Dong Chen$^{2}$,
Fang Wen$^{2}$ }
%
\institute{$^1$University of Science and Technology of China and
$^2$Microsoft Research \\
\email{gsy777@mail.ustc.edu.cn\ and \{jianbao, doch, fangwen\}@microsoft.com}}
\maketitle

\begin{abstract}
Generative adversarial networks (GANs) achieve impressive results today, but not all generated images are perfect. A number of quantitative criteria have recently emerged for generative models, but none of them are designed for a single generated image.
In this paper, we propose a new research topic, Generated Image Quality Assessment (GIQA), which quantitatively evaluates the quality of each generated image. 
We introduce three GIQA algorithms from two perspectives: learning-based and data-based. We evaluate a number of images generated by various recent GAN models on different datasets and demonstrate that they are consistent with human assessments. 
Furthermore, GIQA is available for many applications, like separately evaluating the realism and diversity of generative models, and enabling online hard negative mining (OHEM) in the training of GANs to improve the results. Our code is available now (https://github.com/cientgu/GIQA).

\end{abstract}

\section{Introduction}

Recent studies have shown remarkable success in generative models for their wide applications like high quality image generation~\cite{karras2017progressive,brock2018large}, image-to-image translation~\cite{isola2017image,wang2018high,gu2018arbitrary,gu2019mask}, data augmentation~\cite{frid2018synthetic,frid2018gan}, and so on. However, due to the large variance in quality of generated images, not all generated images are satisfactory for real-world applications. Relying on a manual quality assessment of generated images takes a lot of time and effort. This work proposes a new research topic: Generated Image Quality Assessment (GIQA). The goal of GIQA is to automatically and objectively assess the quality of each image generated by the various generative models.

\begin{figure*}[t]
	\centering
	\includegraphics[width=0.98 \columnwidth]{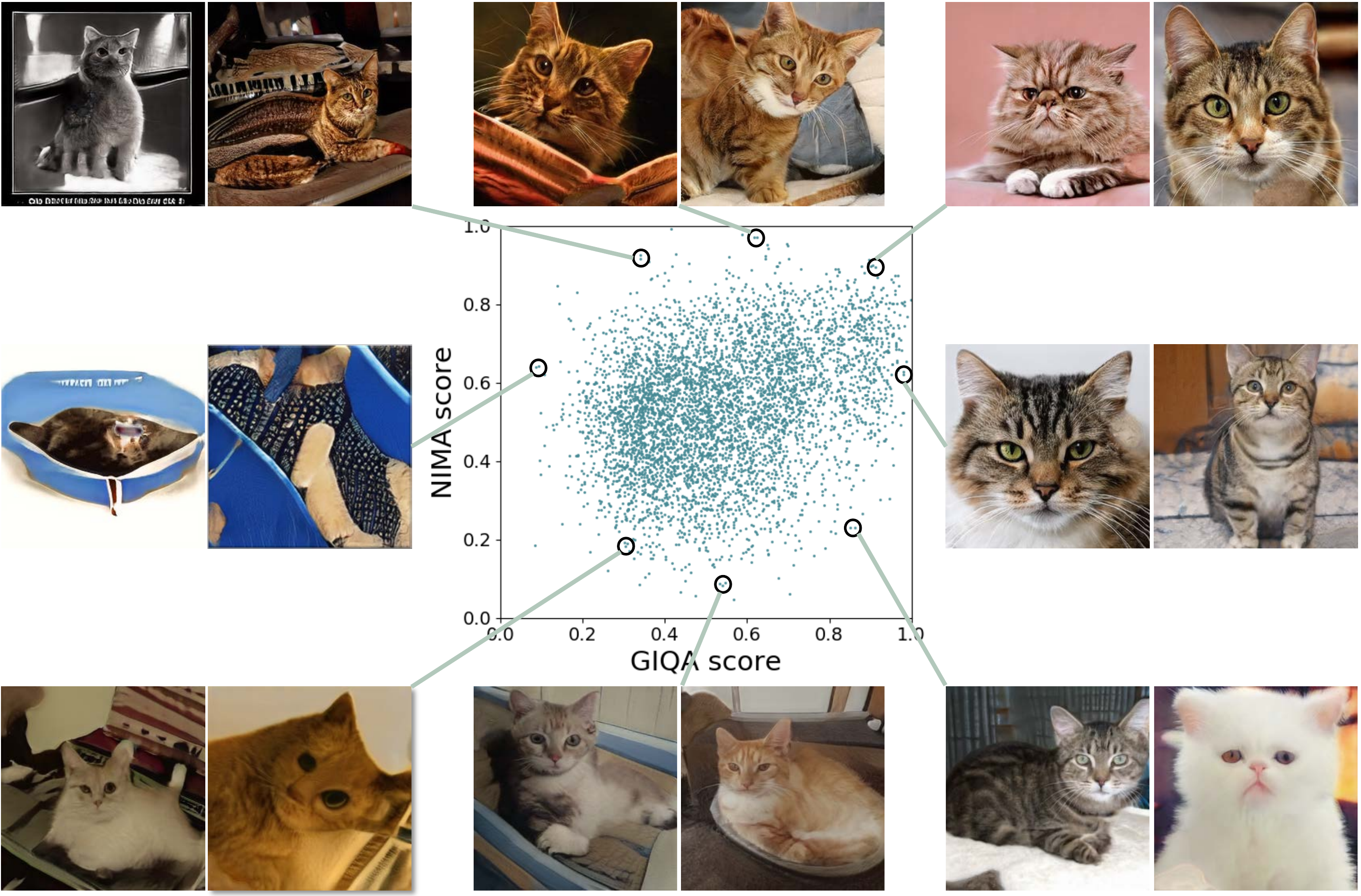}
	\caption{Score distribution of a NR-IQA method NIMA~\cite{talebi2018nima} and our GIQA methods. Score of NIMA is normalized to [0,1] for better comparison, higher score denotes higher image quality. Our GIQA score is more consistent with human evaluation.}
	\label{fig:intro}
	\vspace{-0.7em}
\end{figure*}

GIQA is related to Blind/No-Reference Image Quality Assessment (NR-IQA)~\cite{yan2013no,golestaneh2013no,saad2012blind,moorthy2010two,talebi2018nima}. However, NR-IQA mainly focuses on quality assessment of natural images instead of the generated image. Most of them are distortion-specific; they are capable of performing NR-IQA only if the distortion that afflicts the image is known beforehand, \eg, blur or noise or compression and so on. While the generated images may contain many uncertain model specific artifacts like checkboards~\cite{odena2016deconvolution}, droplet-like~\cite{karras2019analyzing}, and unreasonable structure~\cite{zhang2018self}. Unlike low-level degradations, these artifacts are difficult to simulate at different levels for training. Therefore, traditional natural image quality assessment methods are not suitable for generated images as shown in Figure~\ref{fig:intro}. On the other hand, previous quantitative metrics, like Inception Score~\cite{gulrajani2017improved} and FID~\cite{heusel2017gans}, focus on the assessment of the generative models, which also cannot be used to assess the quality of each single generated image.

In this paper, we introduce three GIQA algorithms from two perspectives: learning-based and data-based perspectives. For the learning-based method, we apply a CNN model to regress the quality score of a generated image. The difficulty with this approach is that different generative models may have their own unique degradation. It is almost impossible to obtain a large amount of manually labelled data to cover all kinds of degradations. Therefore, we propose a novel semi-supervised learning procedure. We observe that the quality of the generated images gets better and better during the training process of generative models. Based on this, we use images generated by models with different iterations, and use the number of iterations as the pseudo label of the quality score. To eliminate the label noise, we propose a new algorithm that uses multiple binary classifiers as a regressor to implement regression. Our learning-based algorithm can be applied to a variety of different models and databases without any manual annotation.


For data-based methods, the essence is that the similarity between the generated image and the real image could indicate quality, so we convert the GIQA problem into density estimation problem of real images. This problem can be broadly categorized as  a parametric and non-parametric method. For the parametric method, we directly adopt the Gaussian Mixture Model (GMM) to capture the probability distribution of real data, then we estimate the probability of a generated image as the quality score. Although this model is very simple, we find it works quite well for most situations. A limitation of the parametric method is that the chosen density might not capture complex distribution, so we propose another non-parametric method by computing the distance between generated image and its K nearest neighbours (KNN), the smaller distance indicates larger probability. 

The learning-based method and the data-based method each have their own advantages and disadvantages. The GMM based method is easy to use and can be trained without any generated images, but it can only be applied to relatively simple distributed databases. The KNN based method has a great merit that there is no training phase, but its memory cost is large since it requires the whole training set to be stored. The learning-based method can handle a variety of complex data distributions, but it is also very time-consuming to collect the images generated by various models at different iterations. Considering both effectiveness and efficiency, we recommend GMM-GIQA mostly. We evaluate these 3 methods in detail in the experiments part.

The proposed GIQA methods can be applied in many applications. 1) We can apply it for generative model assessment. Current generative model assessment algorithms like Inception Score~\cite{gulrajani2017improved} and FID~\cite{heusel2017gans} evaluate the performance of the generative model in a score which represents the summation of two aspects: realism and diversity. Our proposed GIQA model can evaluate these two aspects separately. 2) By using our GIQA method, we can assess the quality of generated images for a specific iteration of generator, and rank the quality of these samples, we suggest that the generator pay more attention to the samples with low quality. To achieve this, we adopt online hard negative mining (OHEM)~\cite{shrivastava2016training} in the discriminator to put larger loss of  weight to the lower quality generated samples. Extensive experiments demonstrate that the performance of the generator is improved by this strategy. 3) We can leverage GIQA as an image picker to obtain a subset of generated images with higher quality.

Evaluating the GIQA algorithm is an open and challenging problem. It is difficult to get the precise quality annotation for the generated images. In order to evaluate the performance of our methods, we propose a labeled generated image for the quality assessment (LGIQA) dataset. To be specific, we present a series of pairs which consist of two generated images for different observers to choose which has a better quality. We keep the pairs which are annotated with the consistent opinions for evaluating. We will release the data and encourage more research to explore the problem of GIQA.

To summarise, our main contribution are as follows:
\begin{enumerate}
	\item To our knowledge, we are the first to propose the topic of generated image quality assessment (GIQA). We proposed three novel methods from two perspectives.
	\item Our method is general and available to many applications, such as separately evaluating the quality and diversity of generative models and improving the results of generative model through OHEM.
	\item We release the LGIQA dataset for evaluating different GIQA methods.
\end{enumerate}

\section{Related Work}
\vspace{-0.08cm}
In this section, we briefly review prior natural image quality assessment methods and generative model assessment methods that are most related to our work.

\noindent \textbf{Image Quality Assessment}:
Traditional Image Quality Assessment(IQA) aims to assess the quality of natural images regarding low-level degradations like noise, blur, compression artifacts, \etc. It is a traditional technique that is widely used in many applications. Formally, it can be divided into three main categories: Full-reference IQA (FR-IQA), Reduced-reference IQA (RR-IQA) and No-reference IQA (NR-IQA). FR-IQA is a relatively simple problem since we have the reference for the image to be assessed, the most widely used metrics are PSNR~\cite{huynh2008scope} and SSIM~\cite{wang2004image}. RR-IQA~\cite{wang2005reduced} aims to evaluate the perceptual quality of a distorted image through partial information of the corresponding reference image. NR-IQA is a more common real-world scenario which needs to estimate the quality of natural image without any reference images. Many NR-IQA approaches~\cite{yan2013no,golestaneh2013no,saad2012blind,moorthy2010two} focus on some specific distortion. Recently advances in convolution neural networks (CNNs) have spawned many CNNs based methods~\cite{kang2014convolutional,kang2015simultaneous,bosse2016deep} for natural image quality assessment. More recent works~\cite{pan2018blind,lin2018hallucinated} leverage the generative model and encourage the model to learn the intermediate perceptual meaning (quality map and hallucinated reference) first and then regress the final quality score.

\noindent \textbf {Generative Model Assessment}:
Recent studies have shown remarkable success in generative models. Many generative models like VAEs~\cite{doersch2016tutorial}, GANs~\cite{goodfellow2014generative}, and Pixel CNNs~\cite{van2016conditional} have been proposed, so the assessment of generative models has received extensive attention. Many works try to evaluate the generative model by conducting the user study, users are often required to score the generated images. While this will cost a large amount of time and effort. Therefore early work~\cite{salimans2016improved} propose a new metric Inception Score (IS) to measure the performance of generative model,  the Inception Score evaluates the generative model in two aspects: realism and diversity of the generated images which are synthesized using the generative model. More recent work~\cite{heusel2017gans} proposes the Fréchet Inception Distance (FID) score for the assessment of generative models. It takes the real data distribution into consideration and calculates the statistics between the generated samples distribution and real data distribution. ~\cite{kynkaanniemi2019improved} proposed precision and recall to measure generative model from quality and diversity separately.

\vspace{-0.1cm}

\section{Methods}

Given a generated image $\mathcal{I}_g$, the target of generated image quality assessment (GIQA) is to quantitatively and objectively evaluate its quality score $S(\mathcal{I}_g)$ which should be consistent with human assessment. We propose solving this problem from two different perspectives. The first one is a learning-based method, in which we apply a CNN model to regress the quality score of a generated image. The second one is a data-based method, for which we directly model the probability distribution of real data. Thus we can estimate the quality of a generated image by the estimated probability from the model. We'll describe them in detail in the following sections.

\subsection{Learning-based Methods}
For learning-based methods, we aim to apply a CNN model to learn the quality of the generated images. Previous supervised learning method often require large amounts of labeled data for training. However, the quality annotation for the generated images is difficult to obtain since it is impossible for human observers to give the precise score to each generated image. Therefore, we propose a novel semi-supervised learning procedure. 

\noindent \textbf{Semi-Supervised learning}: We find an important observation that the quality of generated images from most generative models, \eg,  PGGAN~\cite{karras2017progressive} and StyleGAN~\cite{karras2019style}, is becoming better and better as the training iteration increases. Based on this, we collect images generated by models with different iterations, and use the number of iterations as the pseudo label of the quality score. Note that there is still a gap between the quality of the image generated by the last iteration and the real image. So we suppose that the quality of the generated images ranges from $0$ to $S_{g}$, where $S_{g}\in (0, 1)$, and the quality of the real images is $1$. Formally, the pseudo label of quality score $S_{\text{p}}(\mathcal{I})$ for image $\mathcal{I}$ is
\begin{equation}
\label{eqn:quality_score}
S_{\text{p}}(\mathcal{I}) = 
\begin{cases}
 \frac{S_{g} \cdot iter}{max\_iter} & \textit{if }\mathcal{I}\textit{ is generated}\\
1 & \textit{otherwise}
\end{cases},
\end{equation} 
where $iter$ presents the iteration number, $max\_iter$ presents the maximum iteration number, $S_{g}$ defines the maximum quality score for the generated image, we set it to $0.9$ in our experiment. Then we are able to build a training dataset $\mathcal{D} = \{\mathcal{I}, S_{\text{p}}(\mathcal{I})\}$ for semi-supervised learning, where $\mathcal{I}$ represents the generated images or the real images, $\mathcal{S_{\text{p}}(\mathcal{I})}$ denotes the corresponding quality score.

\noindent \textbf{Multiple Binary Classifiers as Regressor}: A basic solution is to directly adopt a CNN based framework to regress the quality score from the input image. However, we find that this naive regression method is sub-optimal, since the pseudo label contains a lot of noise. Although statistically the longer the training is, the better the quality is, but there is also a large gap in image quality within the same iteration. To solve this problem, inspired by previous work~\cite{liu2016age}, we propose employing multiple binary classifiers to learn the generated image quality assessment, which we call MBC-GIQA. To be specific, $N$ binary classifiers are trained. For the $i$-th classifier,  the training data is divided into positive or negative samples according to a threshold $T^{i}$, given an image $\mathcal{I} \in \mathcal{D}$, its label $c^{i}$ for the $i$-th classifier is: 
\begin{equation}
\label{eqn:quality_score}
c^{i} = 
\begin{cases}
0 & \textit{if }S_p(\mathcal{I}) < T^{i}\\
1 & \textit{otherwise}
\end{cases},
\end{equation} 
where $i = 1, 2, \dots, N$ and $0 < T^{1} < T^{2} < \dots < T^{N} = 1$. So a quality score $S_p(\mathcal{I})$ can be converted to a set of binary labels $\{c^{1},c^{2},\dots, c^{N}\}$. Each binary classifier learns to distinguish whether the quality value is larger than $T^{i}$. Suppose the predicted score for $i$-th binary classifier is $\hat{c}^i$, $i = 1,2,\dots,N$. So the training loss for the framework is:
\begin{align}
L = -\sum_{I \in \mathcal{D}} \sum_{i=1}^{N}(c^i \text{log}(\hat{c}^i) + (1-c^{i})\text{log}(1-\hat{c}^i)).
\end{align}

Using classification instead of regression in this way can be more robust to noise. Although both positive and negative training samples contain noise, $T^i$ is still statistically the decision boundary of $i$-th classifier. During the inference time, suppose we get all the predicted scores $\hat{c}^i, i=1,2,\dots,N$ for a generated image $\mathcal{I}_g$. Then the final predicted quality score for $\mathcal{I}_g$ is the average of all predicted scores: 
\begin{equation}
S_{\text{MBC}}(\mathcal{I}_g) = \frac{1}{N}\sum_{i=1}^{N}{\hat{c}^i}.
\end{equation} 

\begin{figure*}[t]
	\centering
	\includegraphics[width=1.0\columnwidth]{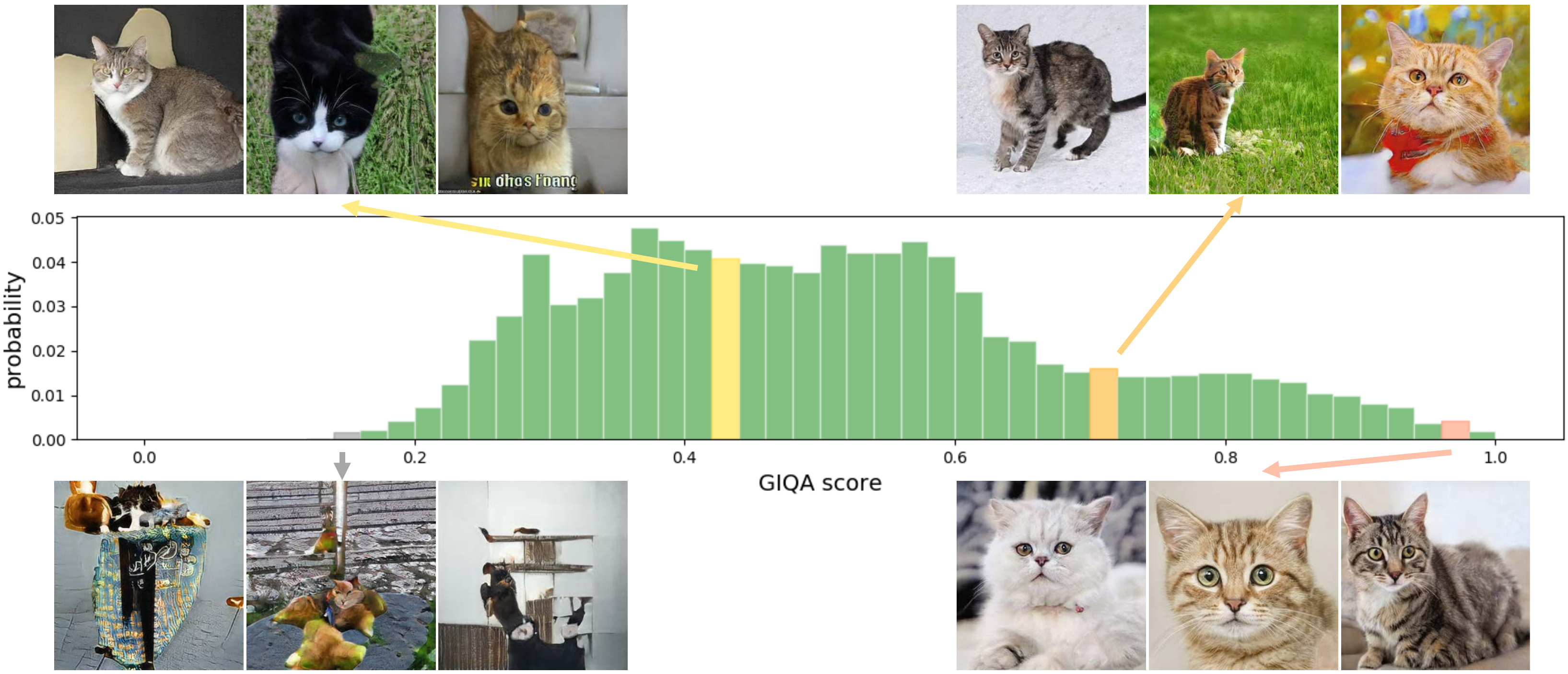}
	\vspace{-0.2cm}
	\caption{Generated images from StyleGAN pretrained on LSUN-cat dataset, sorted by GMM-GIQA method, we random sample images from different score for better visualizion.}
	\label{fig:styleGAN_distribution}
	\vspace{-0.1cm}
\end{figure*}

\vspace{-0.2cm}
\subsection{Data-based Methods}
\vspace{-0.07cm}
Data-based methods aims to solve the quality estimation in a probability distribution perspective. We directly model the probability distribution of the real data, then we can estimate the quality of a generated image by the estimated probability from the model. We propose adopting two density estimation methods: Gaussian Mixture Model (GMM) and $K$ Nearest Neighbour (KNN).

\noindent \textbf{Gaussian Mixture Model}: We propose adopting the Gaussian Mixture Model (GMM) to capture the real data distribution for generated image quality assessment. We call this method GMM-GIQA. A Gaussian mixture model is a weighted sum of $M$ component Gaussian densities. Suppose the mean vector and covariance matrix for $i$-th Gaussian component are $\bm{\mu}^i$ and $\mathbf{\Sigma}^i$, respectively. The probability of an image $\mathcal{I}$ is given by:
\begin{equation}
    p(\mathbf{x}|\lambda) = \sum_{i=1}^{M} \mathbf{w}^i g(\mathbf{x}|\bm{\mu}^i, \mathbf{\Sigma}^i),
    \label{eqn:GMM_probability}
\end{equation}
where $\mathbf{x}$ is the extracted feature of $\mathcal{I}$. Suppose the feature extractor function is $f(\cdot)$, so $\mathbf{x} = f(\mathcal{I})$. $\mathbf{w}^i$ is the mixture weights, which satisfies the constraint that $\sum_{i=1}^{M}\mathbf{w}^i = 1$. And $g(\mathbf{x}|\bm{\mu}^i, \mathbf{\Sigma}^i)$ is the component Gaussian densities. 

The complete Gaussian mixture model is parameterized by the mean vectors, covariance matrices, and mixture weights from all component densities. These parameters are collectively represented by the notation, $\lambda = \{\mathbf{w}^i, \bm{\mu}^i, \mathbf{\Sigma}^i\}$. To estimate these parameters, we adopt the expectation-maximization (EM) algorithm~\cite{dempster1977maximum} to iteratively update them. Since the probability of a generated image represents its quality score, the quality score of $\mathcal{I}_g$ is given by:
\begin{equation}
S_{\text{GMM}}(\mathcal{I}_g) = p(f(\mathcal{I}_g)|\lambda).
\end{equation} 

\noindent \textbf{$K$ Nearest Neighbour}: When the real data distribution becomes complicated, it would be difficult to capture the distribution with GMM well. In this situation, we introduce a non-parametric method based on K Nearest Neighbor (KNN). We think the Euclidean distance between generated images and nearby real images in feature space could also represent the probability of generated image, suppose the feature of a generated sample is $\mathbf{x}$. Its $k$-th nearest real sample's feature is $\mathbf{x}^k$. So we could calculate the probability of generated image as:
\begin{equation}
    p(\mathbf{x}) = \frac{1}{K} \sum_{k=1}^{K} \frac{1}{||\mathbf{x} - \mathbf{x}^k||^2}.
    \label{eqn:nearest_neighbor_distance}
\end{equation}
Suppose the feature extractor function is also $f(\cdot)$, So the quality score of $\mathcal{I}_g$ is given by:
\begin{equation}
S_{\text{KNN}}(\mathcal{I}_g) = p(f(\mathcal{I}_g)).
\end{equation}

Above all, we introduce three approaches to get three forms of quality score function $S(\mathcal{I}_g)$: $S_{\text{MBC}}(\mathcal{I}_g)$, $S_{\text{GMM}}(\mathcal{I}_g)$, and $S_{\text{KNN}}(\mathcal{I}_g)$. We believe these methods will serve as baselines for further research. In terms of recommendation, we recommend the GMM-based method, since this method outperforms other methods(Table~\ref{table:2-hard-choice-results}) and is highly efficient.

\section{Applications}
The proposed GIQA framework is simple and general. In this section, we will show how generated image quality assessment(GIQA) can be applied in many applications, such as generative model evaluation, improving the performance of GANs. 

\subsection{Generative model evaluation}
Generative model evaluation is an important research topic in the vision community. Recently, a lot of quantitative metrics have been developed to assess the performance of a GAN model based on the realism and diversity of generated images, such as Inception Score~\cite{salimans2016improved} and FID~\cite{heusel2017gans}. However, both of them summarise these two aspects. Our GIQA model can separately assess the realism and diversity of generated images. Specifically, we employ the mean quality score from our methods to indicate the realistic performance of the generative model. Supposing the generative model is $G$, the generated samples are $\mathcal{I}_g^i, i=1,2,\dots,N_g$. So the quality score of generator G is calculated with the mean quality of $N_g$ generated samples:

\vspace{-0.3cm}
\begin{equation}
    QS(G) = \frac{1}{N_g}\sum_{i}^{N_g} S(\mathcal{I}_g^i),
    \label{eqn:qs_eqn}
\end{equation}

On the other hand, we can also evaluate the diversity of the generative model $G$. Note that the diversity represents the relative diversity compared to real data distribution. We exchange the positions of real and generated images in data-based GIQA method. \ie, we use generated images to build the model and then evaluate the quality of the real images. 
Considering if the generated samples have similar distribution with real samples, then the quality of the real samples is high. Otherwise, if the generated samples have the problem of "mode collapse", which means a low diversity, then the probability of the real samples become low. This shows by exchanging the position, the quality of real samples is consistent with the diversity of generative models. Suppose the real samples are $\mathcal{I}_r^i, i=1,2,\dots,N_r$, the score function built with generative model $G$ is $S'(\cdot)$. So the diversity score of the generative model is calculated with mean quality of $N_r$ real samples:
\begin{equation}
    DS(G) = \frac{1}{N_r}\sum_{i}^{N_r} S'(\mathcal{I}_r^i),
    \label{eqn:ds_eqn}
\end{equation}

In summary, we have the quality score (QS) and diversity score (DS) to measure the quality and diversity of generative model separately.

\subsection{Improve the performance of GANs}
\label{sec:improve_gans}
Another important application of GIQA is to help the generative model achieve better performance. In general, the quality of generated images from a specific iteration of the generator have large variance, we can assess the quality of these generated samples by using our GIQA method, then we force the generator to pay more attention to these samples with low quality. To achieve this, we employ online hard negative mining (OHEM)~\cite{shrivastava2016training} in the discriminator to apply a higher loss weight to the lower quality samples. To be specific, we set a quality threshold $T_q$. Samples with quality lower than the threshold $T_q$ will be given a large loss weight $w_l > 1$.


\subsection{Image Picker Based on Quality}
Another important application of GIQA is to leverage it as an image picker based on quality. For the wide applications of generative models, picking high quality generated images is of great importance and makes these applications more practical. On the other hand, for a generative model to be evaluated,  we can take full advantage of the image picker to discard these images with low quality to further improve performance.

\section{Experiments}
In this section, we first introduce the overall experiment setups and then present extensive experiment results to demonstrate the superiority of our approach.

\noindent\textbf{Datasets and Training Details} We conduct experiments on a variety of generative models trained on different datasets. For unconditional generative models, we choose WGAN-GP~\cite{gulrajani2017improved}, PGGAN~\cite{karras2017progressive}, and StyleGAN~\cite{karras2019style} trained on FFHQ~\cite{karras2019style}, and LSUN~\cite{yu2015lsun} datasets. For conditional generative models, we choose pix2pix~\cite{isola2017image}, pix2pixHD~\cite{wang2018high}, SPADE~\cite{park2019semantic} trained on Cityscapes~\cite{cordts2016cityscapes} datasets. FFHQ is a large dataset which contains $70000$ high-resolution face images. LSUN contains $10$ scene categories and $20$ object categories, each category contains a large number of images. The Cityscapes dataset is widely used in conditional generative models. In our experiments, we use all the officially released models of these methods for testing.

For learning-based methods, we need the generated images at different iterations of a generative model for training. Specifically, for unconditional generative models, we collect the generated images in the training process of StyleGAN for training, and test the resulting model on the generated images from PGGAN, StyleGAN, and real images. For the conditional generative model, we use the generated images at different iterations of pix2pixHD for training, and test it on the generated images from pix2pix, pix2pixHD, SPADE and real images. To get these training images, we use the official training code, and collect 200,000 generated images, which consist of images from $4000$ iterations, $50$ images per iteration. We adopt $8$ binary classifiers for the MBC-GIQA approach. For the GMM-GIQA method, we set the number of Gaussian components to $7$ for LSUN and Cityscapes datasets, and $70$ for FFHQ. For the KNN-GIQA method, we set $K$ to $1$ for FFHQ and Cityscapes datasets, $3500$ for LSUN. All features are extracted from the inception model~\cite{szegedy2015going} which is trained on ImageNet. More details please refer to the supplementary material.


\noindent\textbf{Evaluation Metrics} Evaluating GIQA algorithms is an open and challenging problem. To quantitatively evaluate the performance of these algorithms, we collect a dataset which is annotated by multiple human observers. To be specific, we first use the generated images from PGGAN, StyleGAN, and real images to build $1500$ image pairs$\footnote{We not only collect images from pretrained models, but also some low quality images from the training procedure.}$, then we demonstrate these pairs to $3$ human observers to choose which image has a better quality. Finally, we discard the pairs which have inconsistent human opinions. The number of remaining pairs are $974$, $1206$, and $1102$ for FFHQ, LSUN-cat, and Cityscapes dataset, respectively. We refer to this dataset as the Labeled Generated Image Quality Assessment(LGIQA) dataset. To evaluate a GIQA algorithm, we employ the algorithm to rank the image quality in each pairs and check if it is consistent with the annotation. Thus we can calculate the accuracy of each algorithm.

\subsection{Comparison with Recent Works}
Since no previous approach aims to solve the problem of GIQA, we design several baselines and compare our approach
with these baselines to prove the superiority of our approach. 

The first baselines are the methods for natural image quality assessment, we choose recent works like DeepIQA~\cite{bosse2016deep}, NIMA~\cite{talebi2018nima}, RankIQA~\cite{liu2017rankiqa} for comparison. For DeepIQA and NIMA, we directly apply their released model for testing. For RankIQA, we use their degradation strategy and follow their setting to train a model on our datasets.  
The second baselines are related to the learning-based method. We adopt the simple idea of directly employing a CNN network to regress pseudo label of quality score $S_p(\mathcal{I})$, which is called IR-GIQA. Another idea is instead of using multiple binary classifiers, we use only 1 classifier to determine whether the image is real or not, we call this BC-GIQA. The third baseline belongs to the data-based method. A simple idea to capture the real data probability distribution is to use a single Gaussian model, we call this SGM-GIQA. 

We present the comparison on LGIQA in Table~\ref{table:2-hard-choice-results}. We observe that our proposed GIQA methods perform better than those natural image assessment methods. Meanwhile, the MBC-GIQA gets higher accuracy than the baseline IR-GIQA and BC-GIQA, and GMM-GIQA is also better than the SGM-GIQA model. which demonstrates the effectiveness of our proposed method. Overall, GMM-GIQA achieves the best results, so we use GMM-GIQA for the following experiments.

 \begin{figure*}[t]
\centering
 \includegraphics[width=1.0\columnwidth]{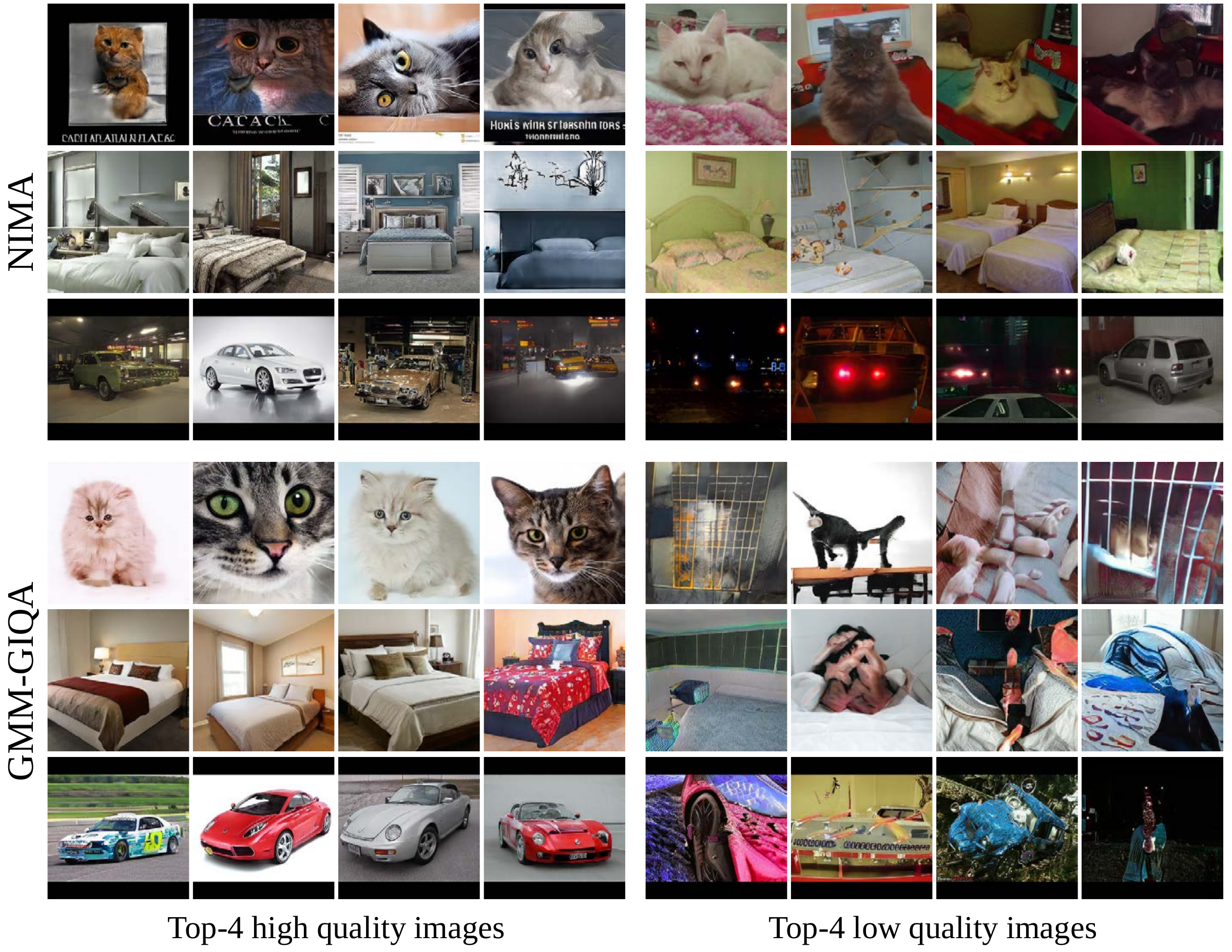}
 \vspace{-0.7cm}
\caption{Generated image quality assessment results for NIMA(the top 3 rows) and our proposed GMM-GIQA(the bottom 3 rows) on LSUN-cat, LSUN-bedroom, and LSUN-car datasets. The left are the top-$4$ high quality images and the right are top-$4$ low quality images.}
\label{fig:select_results}
\vspace{-0.3cm}
\end{figure*}

\begin{table}[t!]
\centering
\begin{tabular}{l|c|c|c}
 \hline
 methods & \ \  FFHQ \ \  &  LSUN-cat  &  Cityscapes \\
 \shline
 NIMA\cite{talebi2018nima} & 0.598  & 0.583   &  0.827 \\
 DeepIQA\cite{bosse2016deep} & 0.581  &  0.550  &  0.763 \\
 RankIQA\cite{liu2017rankiqa} & 0.573  &  0.557  &  0.780 \\
 \hline
 BC-GIQA & 0.663  & 0.710   & 0.768  \\ 
 IR-GIQA & 0.678  &  0.784  &  0.837 \\ 
 SGM-GIQA & 0.620  & 0.829 & 0.847 \\
 \hline
  MBC-GIQA(\textbf{our}) & 0.731  &  0.831  &  0.886 \\
 GMM-GIQA(\textbf{our})  & \textbf{0.764}  &  \textbf{0.846}  & 0.895  \\ 
 KNN-GIQA(\textbf{our})  & 0.761  &  0.843  & \textbf{0.898}  \\ 
  \hline
\end{tabular}
\vspace{0.10cm}
\caption{Comparison of the accuracy on LGIQA dataset for different methods.}
\label{table:2-hard-choice-results}
\vspace{-1.1cm}
\end{table} 

We qualitatively compare the generated image quality ranking results for our proposed GMM-GIQA and NIMA in Fig~\ref{fig:select_results}, we can observe that GMM-GIQA achieve a better generated image quality ranking results that is more consistent with human assessment. More results can be found in supplemental material.

\vspace{-0.2cm}

\subsection{Generative Model Assessment}

\begin{figure*}[t]
\centering
 \includegraphics[width=1.0\columnwidth]{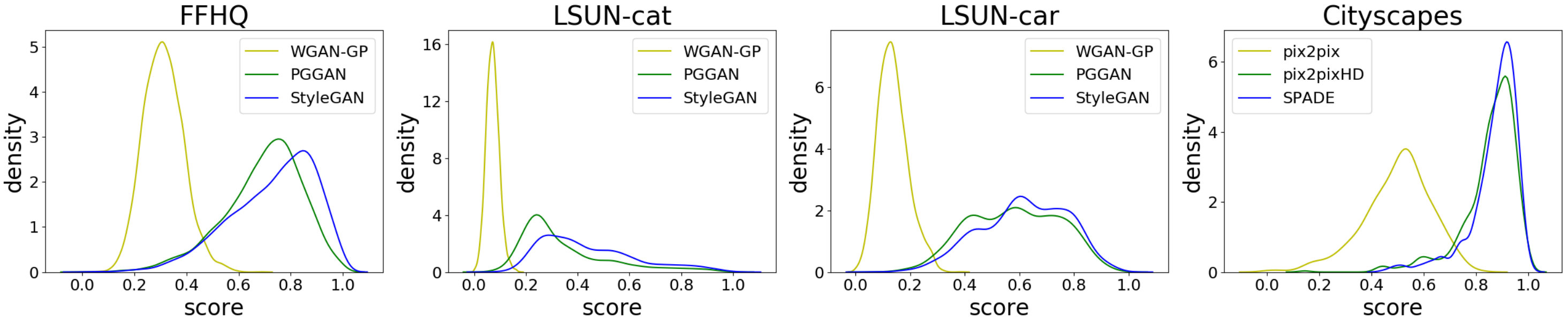}
 \vspace{-0.7cm}
\caption{Quality score distribution of generated images from different generative models.}
\label{fig:model_distribution}
\vspace{-0.2em}
\end{figure*}

\noindent \textbf{Quality Distribution Evaluation} The proposed GIQA methods are able to assess the quality for every generated sample. Therefore we first employ our proposed GMM-GIQA to validate the quality distribution of generated samples from several generative models. For unconditional generative models, we choose WGAN-GP, PGGAN, StyleGAN trained on FFHQ, LSUN-cat and LSUN-car datasets. For conditional generative models, we choose pix2pix, pix2pixHD, and SPADE trained on Cityscapes dataset. Each generative model generates $5000$ test images, and then apply our GMM-GIQA method to calculate the quality score, the quality score distributions are shown in Figure~\ref{fig:model_distribution}. Note that all the quality scores are normalized to $[0 ,1]$. We find that PGGAN and StyleGAN are much better than WGAN-GP, and StyleGAN is better than PGGAN. SPADE and pix2pixHD are much better than pix2pix, SPADE is slightly better than pix2pixHD. All these observations are consistent with human evaluation.

\begin{table}[t!]
\scriptsize
\centering
  \begin{tabular}{c|ccc|ccc|ccc}
    \hline
    \multirow{2}{*}{ } &
      \multicolumn{3}{c|}{FFHQ } &   
      \multicolumn{3}{c|}{LSUN-cat } &  
      \multicolumn{3}{c}{LSUN-car } \\
    \cline{2-10}
    &  WGAN-GP  &  PGGAN  & StyleGAN & WGAN-GP & PGGAN & StyleGAN  & WGAN-GP & PGGAN & StyleGAN \\
    \shline
    FID & 107.6 & 14.66 & 10.54 & 192.2 & 49.87 & 18.67 & 146.5 & 14.73 & 12.70 \\
    Prec & 0.006 & 0.640 & 0.704 & 0.012 & 0.487 & 0.608 & 0.022 & 0.608 & 0.680 \\
    Rec & 0 & 0.452 & 0.555 & 0 & 0.356 & 0.467 & 0.002 & 0.487 & 0.531 \\
    QS & 0.312 & 0.694 & 0.731 & 0.072 & 0.347 & 0.441 & 0.138 & 0.583 & 0.617 \\
    DS & 0.355 & 0.815 & 0.806 & 0.236 & 0.789 & 0.796 & 0.281 & 0.801 & 0.791 \\
    \hline
  \end{tabular}
  \vspace{0.14cm}
\caption{Comparison of FID, Precision~\cite{kynkaanniemi2019improved}, Recall~\cite{kynkaanniemi2019improved}, QS, and DS metric for the generative model WGAN-GP, PGGAN, and StyleGAN on three different datasets: FFHQ, LSUN-cat, and LSUN-car.}
\label{table:model_score}
\vspace{-0.3cm}
\end{table}

\begin{table}[t!]
\centering
  \begin{tabular}{l|c|c|c}
    \hline
        & pix2pix \ & pix2pixHD \ & SPADE \\
    \shline
    QS & 0.498 & 0.851 & 0.870 \\
    \hline
  \end{tabular}
  \vspace{0.1cm}
\caption{Quality Score(QS) for pix2pix, pix2pixHD, and SPADE on Cityscapes dataset.}
\label{table:cgan_score}
\vspace{-0.7cm}
\end{table}


\noindent \textbf{QS and DS for Generative Models} As we introduced in Section~\ref{sec:improve_gans}, we propose two new metrics the quality score (QS) and diversity score (DS) to measure the performance of generative models. So we employ these two metrics and FID to quantitatively evaluate these generative models. We use $5000$ generated images to evaluate the FID and quality score (QS), $20000$ generated images and $5000$ real images to evaluate the diversity score (DS). Table~\ref{table:model_score} reports the results on WGAN-GP, PGGAN, and StyleGAN. We observe that our QS metric is consistent with human evaluation, and our DS metric demonstrates that StyleGAN and PGGAN have a better diversity than WGAN-GP. Table~\ref{table:cgan_score} reports the QS metric results for conditional generative models: pix2pix, pix2pixHD and SPADE, we can observe that the results are also consistent with human evaluation.


\begin{table}[t]
\centering
\begin{tabular}{c|c|cccccc}
\hline
 Methods & Metrics & \ \ \ \ 1 \ \ \ \ & \ \ \ \ 0.9 \ \ \ \ & \ \ \ \  0.8 \ \ \ \ & \ \ \ \  0.7 \ \ \ \ & \ \ \ \  0.6 \ \ \ \ & \ 0.5 \\ 
\shline
\multirow{3}{*}{truncation rate}
  & FID & 18.67 & 18.05 & 19.64 & 23.46 & 30.48 & 41.68 \\
  & QS & 0.441 & 0.463 & 0.486 & 0.510 & 0.537 & 0.567 \\
  & DS & 0.796 & 0.771 & 0.756 & 0.731 & 0.699 & 0.686 \\
\hline
\multirow{3}{*}{remaining rate}
  & FID & 18.67 & 16.65 & 17.63 & 20.73 & 25.84 & 33.19 \\
  & QS & 0.441 & 0.466 & 0.495 & 0.520 & 0.551 & 0.587 \\
  & DS & 0.796 & 0.792 & 0.780 & 0.766 & 0.746 & 0.712 \\
\hline
\end{tabular}
\vspace{0.1cm}
\caption{Comparison of truncation trick and image picker trick using StyleGAN on LSUN-cat dataset.}
\vspace{-0.5cm}
\label{table:trick_table}
\end{table}

\begin{table}[t]
\centering
\begin{tabular}{c|c|ccc}
\hline
 Datasets & Methods & \ \ \ FID \ \ \ & \ \ \ \ QS \ \ \ \ \ \ &  DS  \\ 
\shline
\multirow{3}{*}{FFHQ}
  & StyleGAN & 17.35 & 0.697 & 0.753 \\
  & StyleGAN+OHEM & 16.89 & 0.711 & 0.755 \\
  & StyleGAN+OHEM+Picker & 16.68 & 0.723 & 0.749 \\
\hline
\multirow{3}{*}{LSUN-cat}
  & StyleGAN & 18.67 & 0.441 & 0.796 \\
  & StyleGAN+OHEM & 18.12 & 0.462 & 0.790 \\
  & StyleGAN+OHEM+Picker & 16.25 & 0.482 & 0.785 \\
\hline
\end{tabular}
\vspace{0.1cm}
\caption{Performance comparison of various settings for StyleGAN.}
\vspace{-0.8cm}
\label{table:improvement}
\end{table}

\subsection{Improving the Performance of GANs}
One important application of GIQA is to improve the performance of GANs. We find that we can achieve this in two ways, one is to adopt the GIQA to discard low quality images from all the generated images for evaluation. The other one is to take full advantage of the GIQA to achieve OHEM in the training process of GANs, then the performance gets improved. 

\noindent \textbf{Image Picker Trick}
We conduct this experiment on the StyleGAN model trained on LSUN-cat. We first generate $10000$ images, then we use the GMM-GIQA method to rank the quality of these images and retain different percentages of high quality images, finally we randomly sample $5000$ remaining images for evaluation. We test the generated images with different remaining rates. For comparison, we notice that StyleGAN adopt a "truncation trick" on the latent space which also discards low quality images. With a smaller truncation rate, the quality becomes higher, the diversity becomes lower. We report the FID, QS, DS results of different truncation rate and remaining rate in Table~\ref{table:trick_table}. We notice that the FID improves when the truncation rate and remaining rate are set to $0.9$, and the remaining rate works better than the truncation rate. 
Which also perfectly validates the superiority of the QS and DS metric.

\noindent \textbf{OHEM for GANs} To validate whether OHEM improves the performance of GANs, we train two different settings of StyleGAN on FFHQ and LSUN-cat datasets at $256 \times 256$ resolution. One follows the original training setting (denoted as StyleGAN), the other applies the OHEM in the training process and puts a large loss weight $w_l$ on low quality images whose quality score is lower than threshold $T_q$, which we called StyleGAN+OHEM. We set the $w_l$, $T_q$ to $2$, $0.2$ in our experiments. After finishing the training process, we evaluate the FID, QS, and DS metric. Table~\ref{table:improvement} reports the results. We find that OHEM improves the performance of GANs. Besides, based on this model, by using our image picker trick (denoted as StyleGAN+OHEM+Picker), it can further achieve better performance.

\subsection{Analysis of the Proposed Methods}
In this subsection, we conduct experiments to investigate the sensitiveness of hyper parameters in the proposed three approaches. All the results are evaluated on our LGIQA dataset.


\noindent \textbf{Hyper parameters for MBC-GIQA}
For MBC-GIQA, the first parameter we want to explore is how patch size of the training image affects the results. To explore this, we train our multiple binary classifiers on images with different patch size. The training patch size is set to $32$, $64$, $128$, $192$, $256$. During the inference time, we first randomly crop $3$ patches on the test image with the training patch size and then input them to the model to get an average score as the final score. The default number of binary classifiers is set to $8$. We test it on LGIQA-FFHQ dataset, results are shown in Table~\ref{table:patch-size}. We can observe that training images at patch size $192$ gets the best results. Large patch size may lead to bad results, it may be caused by the overfitting problem. Small patch size also leads to bad performance, this may be because small patches can not provide discriminative information to learn the quality assessment. 

The second parameter we want to explore is how the number of binary classifiers influence the results, we train the model using different numbers of binary classifiers. The number of classifiers is set to $1$, $4$, $6$, $8$, $10$, $12$. The training patch size is set to $192$. Table~\ref{table:number-of-bins} reports the results. We find that as the number of binary classifiers increases from $1$ to $8$, the performance becomes better and better, and as the number continues to increase to $12$ the performance degrades.

\begin{table}[t]
\centering
\begin{tabular}{l|c|c|c|c|c|c}
 \hline
 Patch size  &  32  & 64 & 128 & 192 & 224 & 256 \\
 \shline
 Accuracy & 0.686  & 0.709 & 0.725 & \textbf{0.731} & 0.721 & 0.690 \\
 \hline
\end{tabular}
\vspace{0.1cm}
\caption{Results of different training patch size for MBC-GIQA.}
\label{table:patch-size}
\vspace{-0.5cm}
\end{table}

\begin{table}[t!]
\centering
\begin{tabular}{c|c|c|c|c|c|c}
 \hline
 $N$ &  1  & 4 & 6 & 8 & 10 & 12 \\
 \shline
 Accuracy & 0.663  & 0.682 & 0.722 & \textbf{0.731} & 0.718 & 0.717 \\
 \hline
\end{tabular}
\vspace{0.1cm}
\caption{Results of different number of binary classifiers $N$ for MBC-GIQA.}
\label{table:number-of-bins}
\vspace{-0.5cm}
\end{table}

\begin{table}[t!]
\centering
\begin{tabular}{c|c|c|c|c|c|c|c}
 \hline
 $M$ & 5 &  10  &  20  &  30  &  50  &  70 & 100 \\
 \shline
 Accuracy  & 0.648 & 0.663 & 0.738 & 0.733 & 0.752 & \textbf{0.764} & 0.753 \\
 \hline
\end{tabular}
\vspace{0.1cm}
\caption{Results of different number of Gaussian components $M$ for GMM-GIQA.}
\label{table:kernel-number}
\vspace{-0.5cm}
\end{table}

\begin{table}[t!]
\centering
\begin{tabular}{c|c|c|c|c|c|c|c|c|c}
 \hline
 $K$ &  1  & 30 & 100 & 500 & 1000 & 2000 & 3500 & 5000 & 7000 \\
 \shline
 Accuracy  & 0.823   & 0.828 & 0.833 & 0.837 & 0.840 & 0.842 & \textbf{0.843} & 0.841 & 0.840 \\
  \hline
\end{tabular}
\vspace{0.1cm}
\caption{Results of different number of nearest neighbors $K$ for KNN-GIQA.}
\label{table:KNN-abla}
\vspace{-0.5cm}
\end{table}

\noindent \textbf{Hyper parameters of GMM-GIQA}
The key factor for GMM is the number of Gaussian components $M$, therefore we explore how $M$ affects the results of GIQA. We set $M$ to $5$, $10$, $20$, $30$, $50$, $70$, $100$ and test the results on our LGIQA-FFHQ dataset. We show the results in Table~\ref{table:kernel-number}, as the number of Gaussian components $M$ increases from $5$ to $70$, we get better and better results, and the number continues to increase to $100$, the performance degrades.    

\noindent \textbf{Hyper parameters of KNN-GIQA}
To explore how the number of nearest neighbours $K$ affects the results, we apply different $K$ in the KNN-GIQA. Specifically, we set $K$ to $1$, $30$, $100$, $500$, $1000$, $2000$, $3500$, $5000$, $7000$. The result on LGIQA-LSUN-cat are shown in Table~\ref{table:KNN-abla}, as $K$ increases from $1$ to $1000$, we get better and better results, and the number continues to increase to $7000$, the performance is comparable.  

\section{Conclusions}
In this paper, we aim to solve the problem of quality evaluation of a single generated image and propose the new research topic: Generated Image Quality Assessment (GIQA). To tackle this problem, we propose three novel approaches from two perspectives: learning-based and data-based. 
Extensive experiments show that our proposed methods can perform quite well on this new topic, also we demonstrate that GIQA can be applied in a wide range of applications. 

We are also aware that there exist some limitations of our methods. For the learning-based method MBC-GIQA, it requires the generated images at different iterations for training, while these images may not be easily obtained in some situations. For the data-based method GMM-GIQA, there is a chance of failure when the real data distribution is too complicated. We also notice that our current results are far from solving this problem completely. We hope our approach will serve as a solid baseline and help support future research in GIQA.


\clearpage

\bibliographystyle{splncs}
\bibliography{egbib}

\clearpage

\appendix
\section{More details for experiments}
In this section, we provide more training details for MBC-GIQA and GMM-GIQA. For MBC-GIQA, we adopt VGG-19~\cite{simonyan2014very} as the backbone network, only the last layer is replaced with a fully-connected layer with $N$ outputs. $N$ is the number of binary classifiers. We set it to $8$ in our experiments. In the training stage, we split all the $200000$ images into training and validation set with a ratio of $9:1$. We set the batch size to $512$ and total training epoch to $100$, respectively. The learning rate is set to $0.05$ at the beginning of the training and divided by 10 every $25$ epochs. We choose the best performing model based on validation accuracy.

For data-based methods, we use the feature in the final average pooling layer of the InceptionV3 model by default. For GMM-GIQA, we build the Gaussian mixture model based on the scikit-learn platform~\cite{pedregosa2011scikit}. The covariance matrices of the model can be constrained to $4$ types: spherical, diagonal, tied or full covariance. We adopt the full covariance matrices during our experiments.

\section{Examples from LGIQA dataset}
\begin{figure*}[h]
\centering
 \includegraphics[width=1.0\columnwidth]{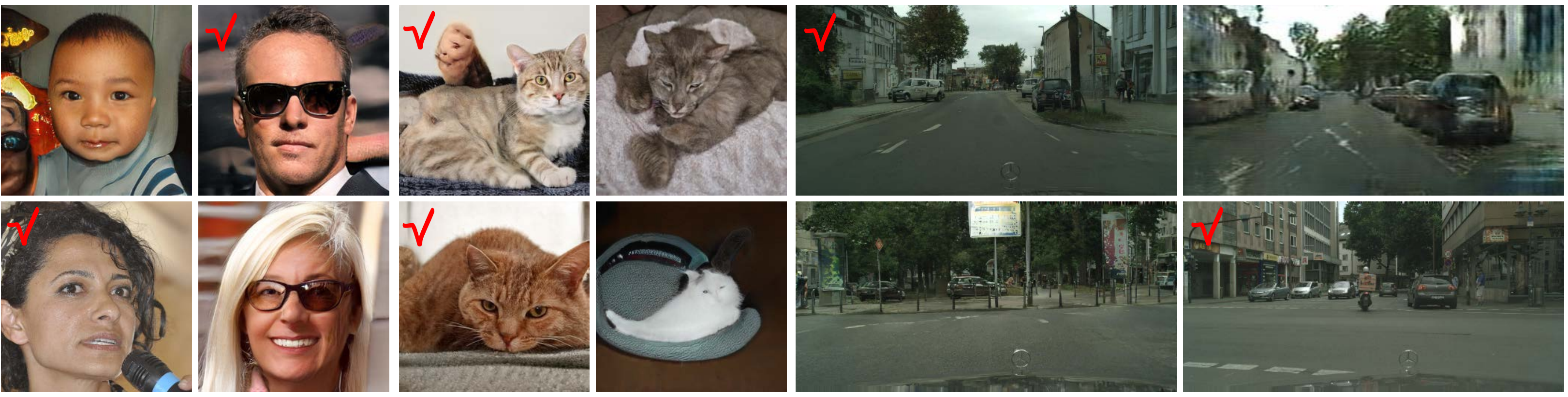}
 \vspace{-0.7cm}
\caption{Some example image pairs from our LGIQA dataset.}
\label{fig:LGIQA}
\vspace{-0.2cm}
\end{figure*}

We propose LGIQA dataset to evaluate the performance of different GIQA methods, it contains generated and real images of $3$ classes: faces, cats and street photos. It consists of image pairs with human annotations of which image has a relatively better quality. We show some example image pairs from our dataset on different classes in Fig~\ref{fig:LGIQA}, we will release this dataset.

\section{More analysis of hyper parameters}
In this section, we present two extra experiments to demonstrate how feature dimensions and the type of covariance matrices affect the performance of the proposed GMM-GIQA method.

\noindent \textbf{Reducing feature dimension for GMM-GIQA}
Since the high feature dimension will cost a large calculation for covariance matrices. We use the traditional method PCA to reduce the dimension of extracted features for GMM. We try to reduce feature dimension to a different percentage of variance. Table~\ref{table:kernel-number} reports the results on LGIQA-FFHQ dataset. We find that although lower dimensions would increase the training speed of GMM, it would lead to a decrease in accuracy. So we suggest not using PCA in GMM-GIQA.

\begin{table}[h]
\centering
\begin{tabular}{c|c|c|c|c}
 \hline
 Percentage of variance  & 0.90 &  0.98  &  0.98  &  1.0   \\
 \shline
  Accuracy & 0.674 & 0.725& 0.755 & \textbf{0.764} \\
 \hline
\end{tabular}
\vspace{0.1cm}
\caption{Results of using PCA to reduce the feature dimension to different percentage of variance for GMM-GIQA.}
\label{table:kernel-number}
\vspace{-0.5cm}
\end{table}

\noindent \textbf{Covariance matrices type of GMM-GIQA}
Another way to reduce the computation cost of covariance matrices is to change the matrices type. We set the covariance matrices with $4$ different types: spherical, diagonal, tied or full covariance, which are donated as f, t, d, s, respectively. we test the resulting model on LGIQA-FFHQ dataset. Table~\ref{table:convariance-matrix} reports the results. We observe that full covariance can get the best results.


\begin{table}[ht]
\centering
\begin{tabular}{c|c|c|c|c}
  \hline
  Types of covariance matrices & f & t & d & s \\
  \shline
  Accuracy & \textbf{0.753} & 0.641 & 0.619 & 0.597  \\
  \hline
\end{tabular}
\caption{Results on using different types of covariance matrices in GMM-GIQA.}
\label{table:convariance-matrix}
\end{table}

\section{More generated image quality assessment results}
\begin{figure*}[h]
\centering
 \includegraphics[width=1.0\columnwidth]{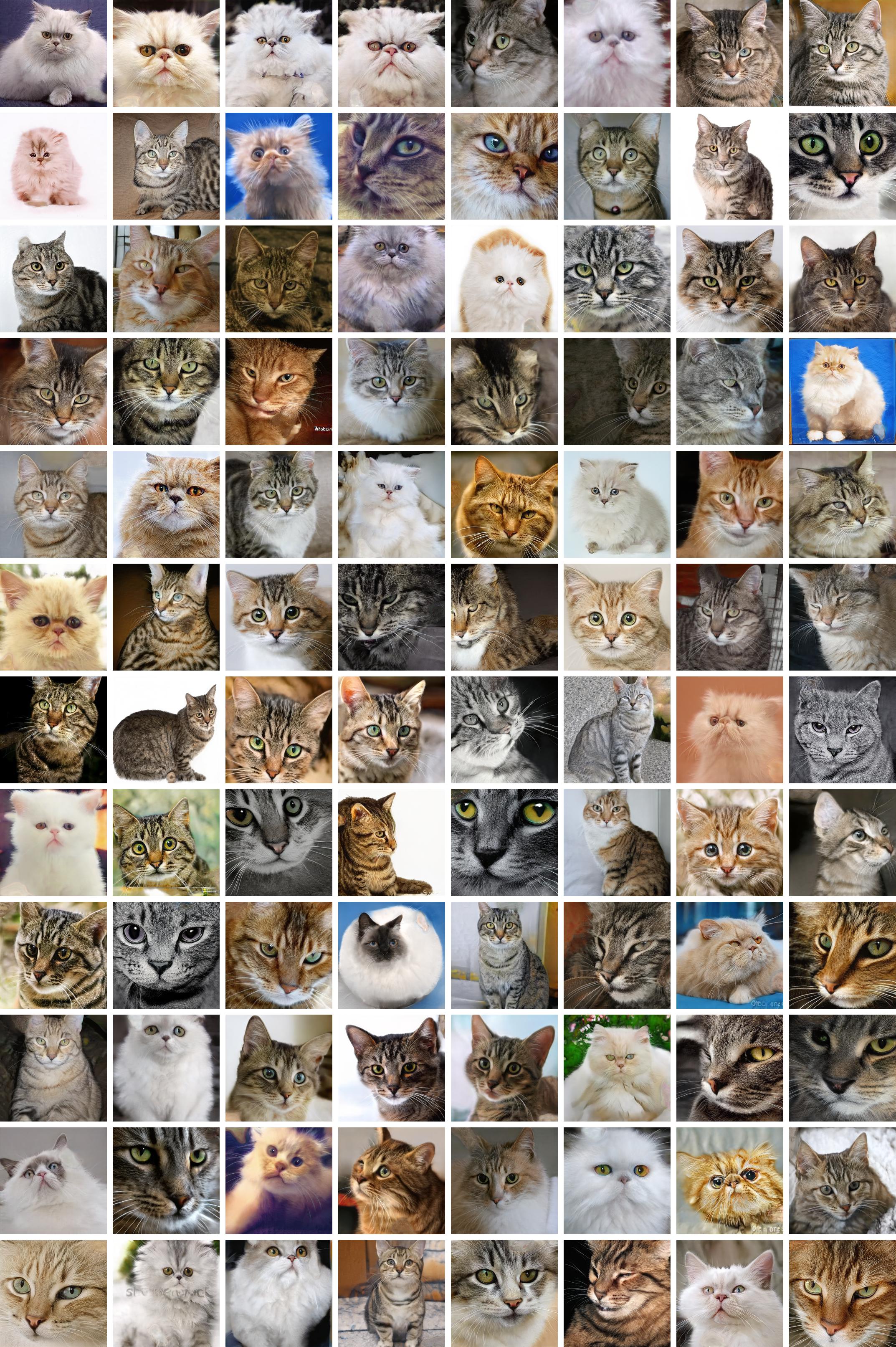}
 \vspace{-0.7cm}
\caption{The top-$96$ high quality generated images from StyleGAN trained on LSUN-cat.}
\label{fig:cat}
\vspace{-0.2cm}
\end{figure*}

\begin{figure*}[h]
\centering
 \includegraphics[width=1.0\columnwidth]{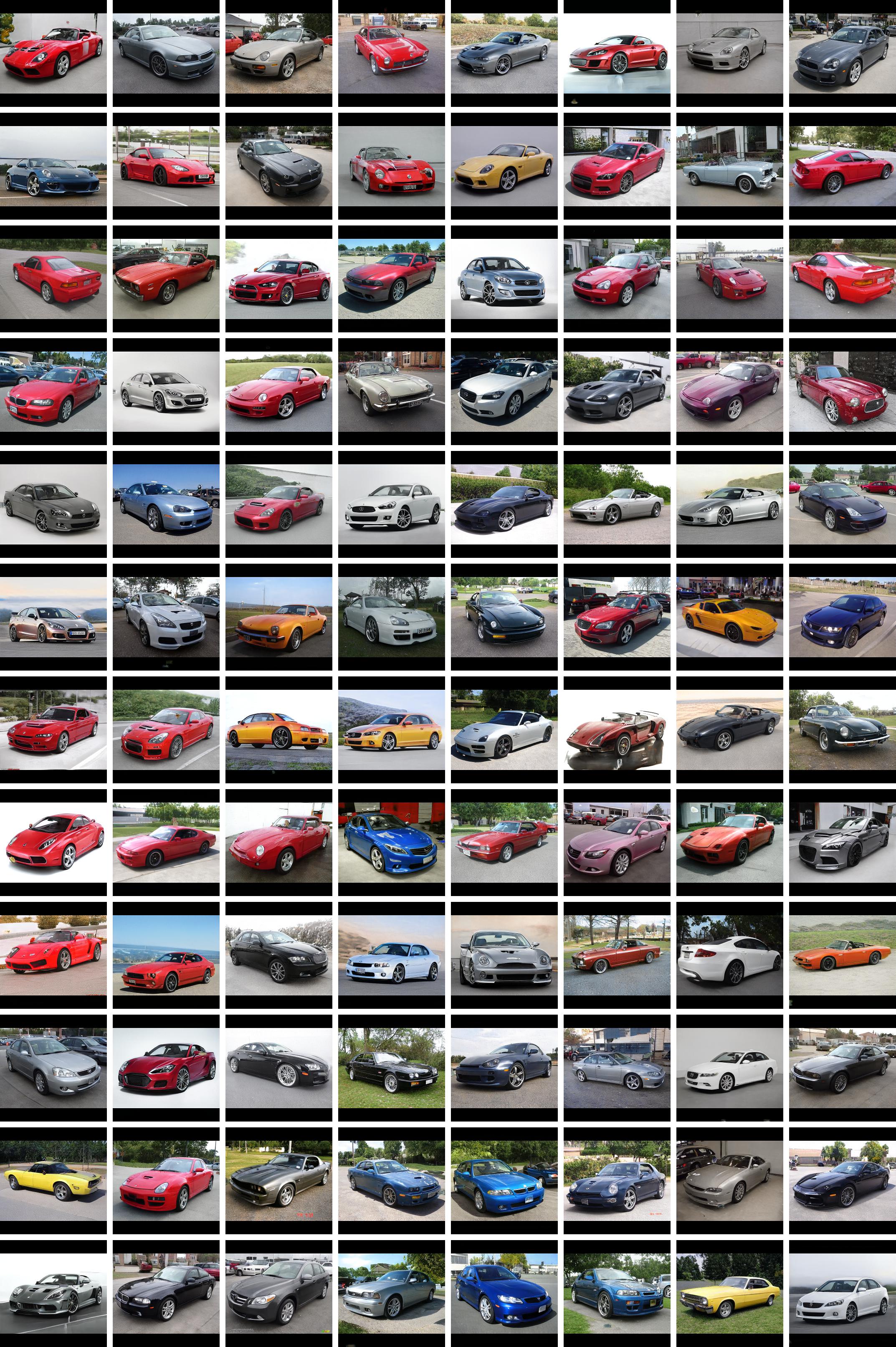}
 \vspace{-0.7cm}
\caption{The top-$96$ high quality generated images from StyleGAN trained on LSUN-car.}
\label{fig:car}
\vspace{-0.2cm}
\end{figure*}

\begin{figure*}[h]
\centering
 \includegraphics[width=1.0\columnwidth]{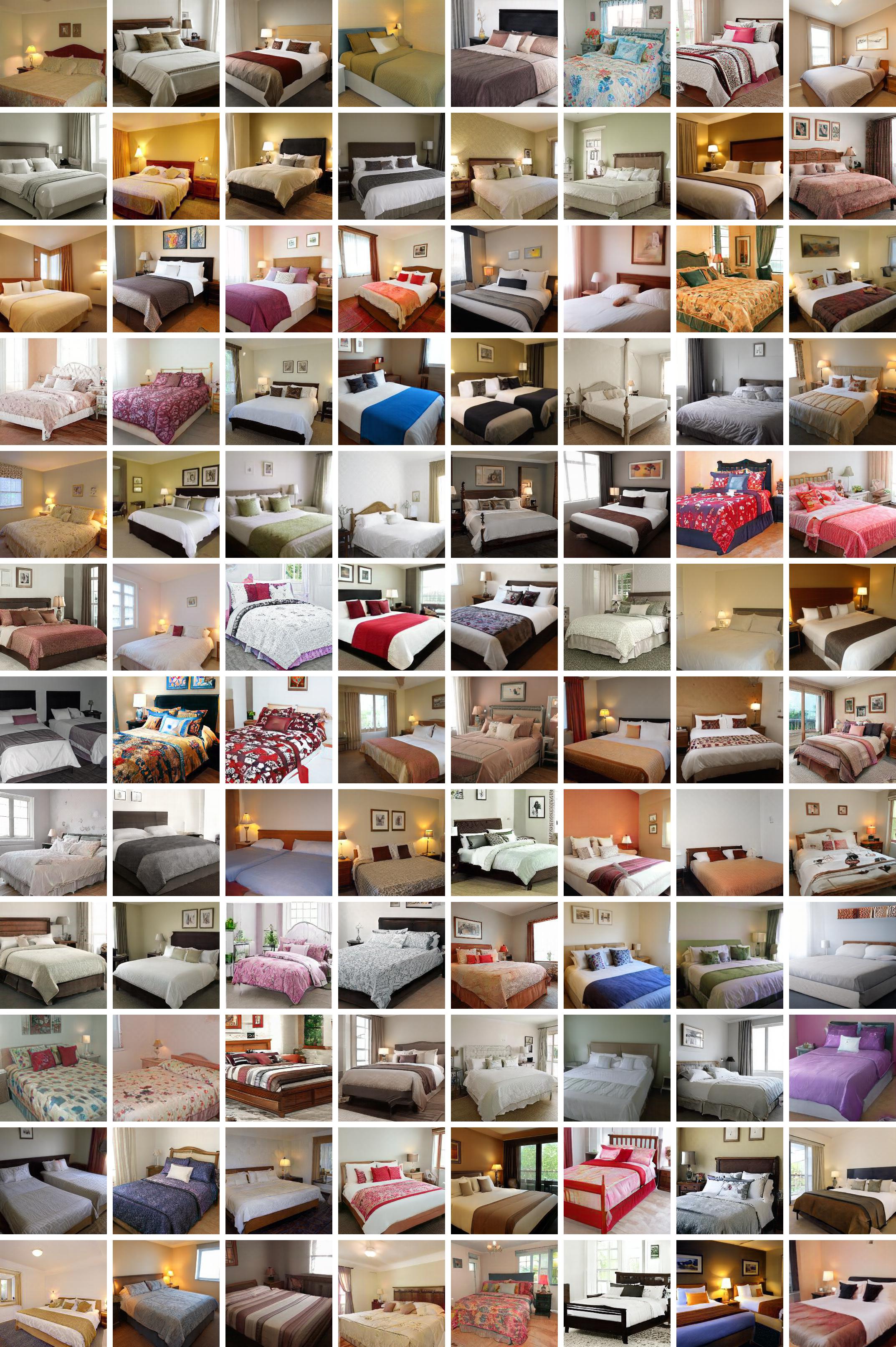}
 \vspace{-0.7cm}
\caption{The top-$96$ high quality generated images from StyleGAN trained on LSUN-bedroom.}
\label{fig:bedroom}
\vspace{-0.2cm}
\end{figure*}

\begin{figure*}[h]
\centering
 \includegraphics[width=1.0\columnwidth]{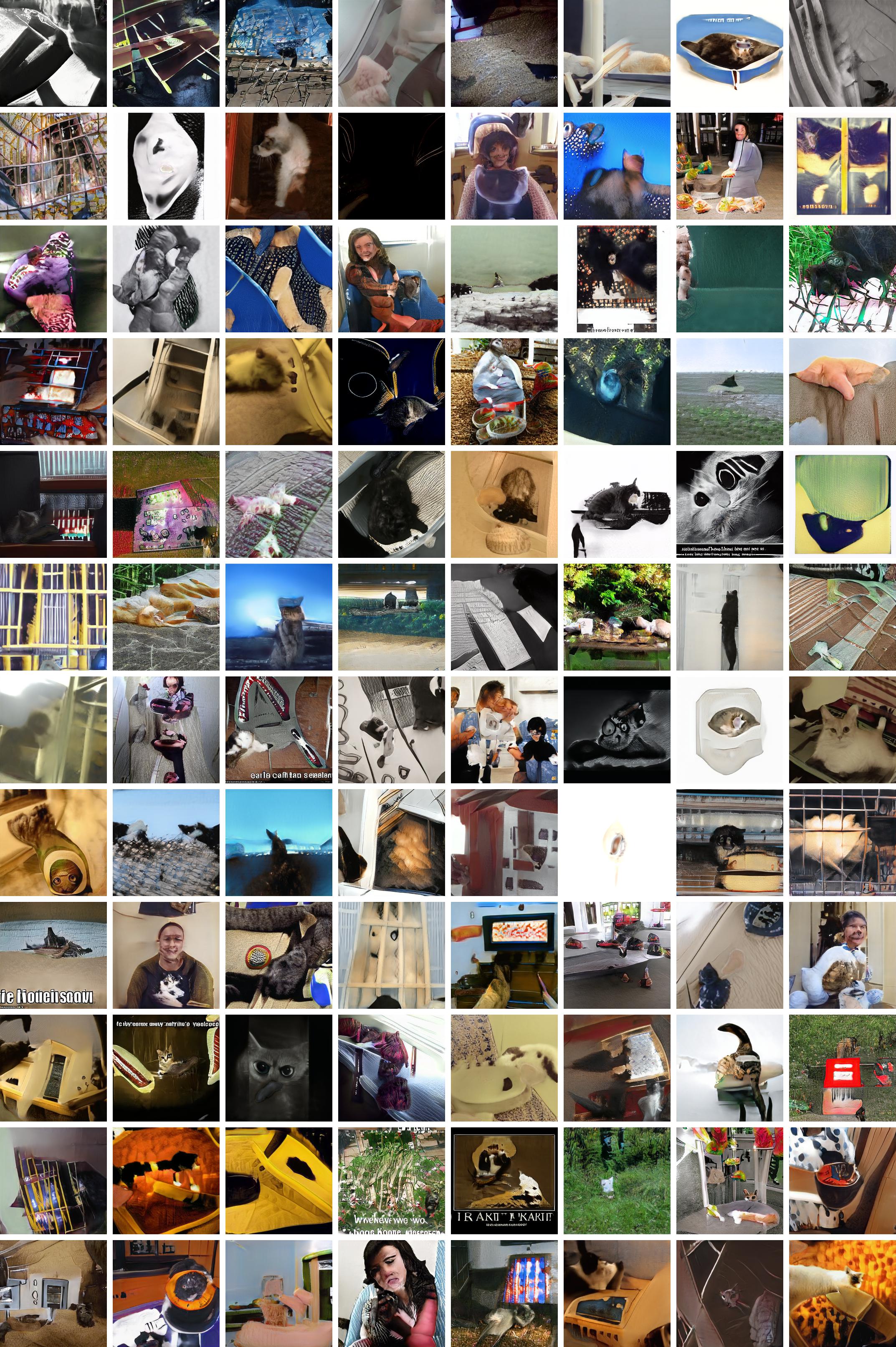}
 \vspace{-0.7cm}
\caption{The top-$96$ low quality generated images from StyleGAN trained on LSUN-cat.}
\label{fig:cat_worst}
\vspace{-0.2cm}
\end{figure*}

\begin{figure*}[h]
\centering
 \includegraphics[width=1.0\columnwidth]{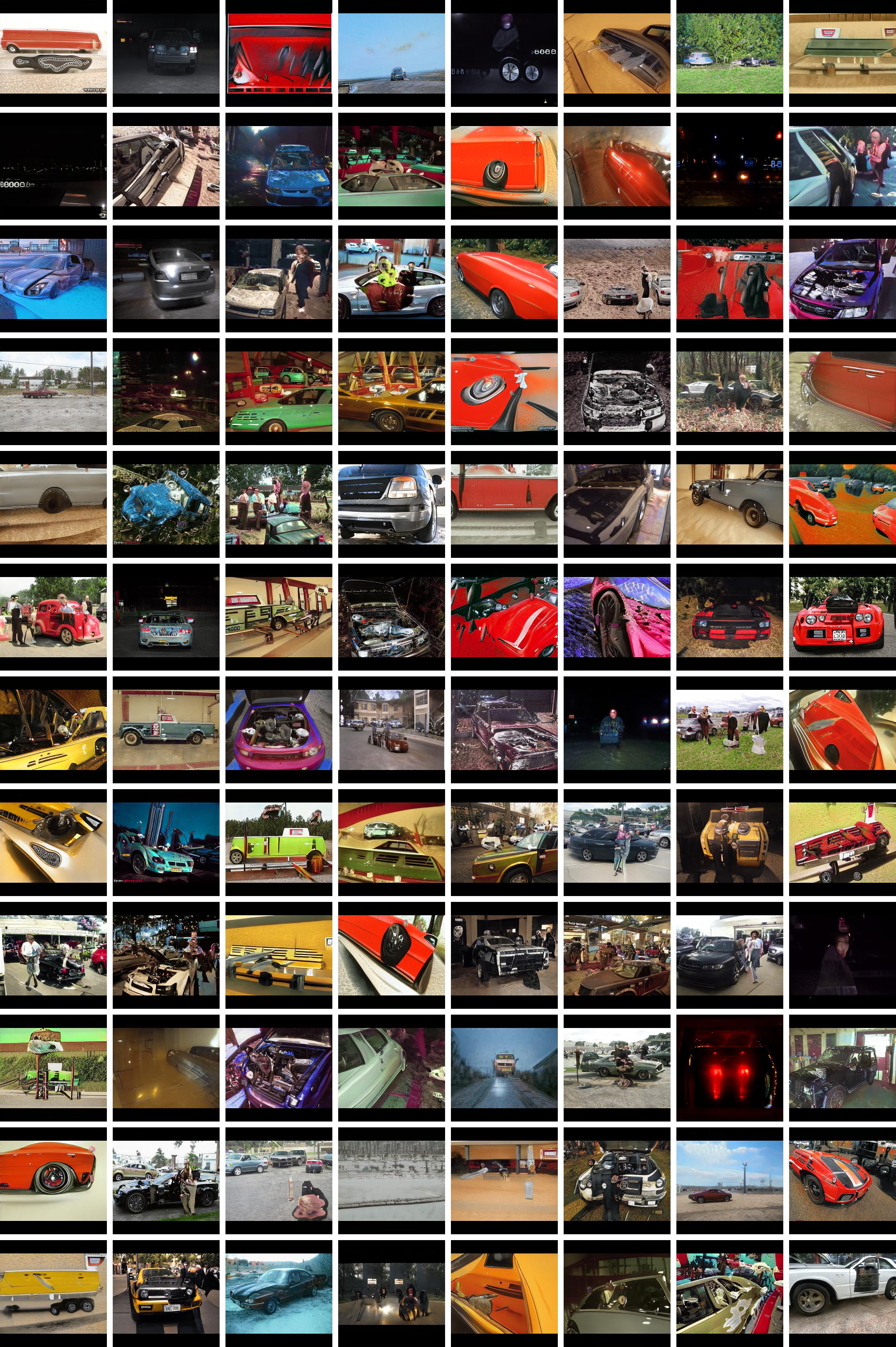}
 \vspace{-0.7cm}
\caption{The top-$96$ low quality generated images from StyleGAN trained on LSUN-car.}
\label{fig:car_worst}
\vspace{-0.2cm}
\end{figure*}

\begin{figure*}[h]
\centering
 \includegraphics[width=1.0\columnwidth]{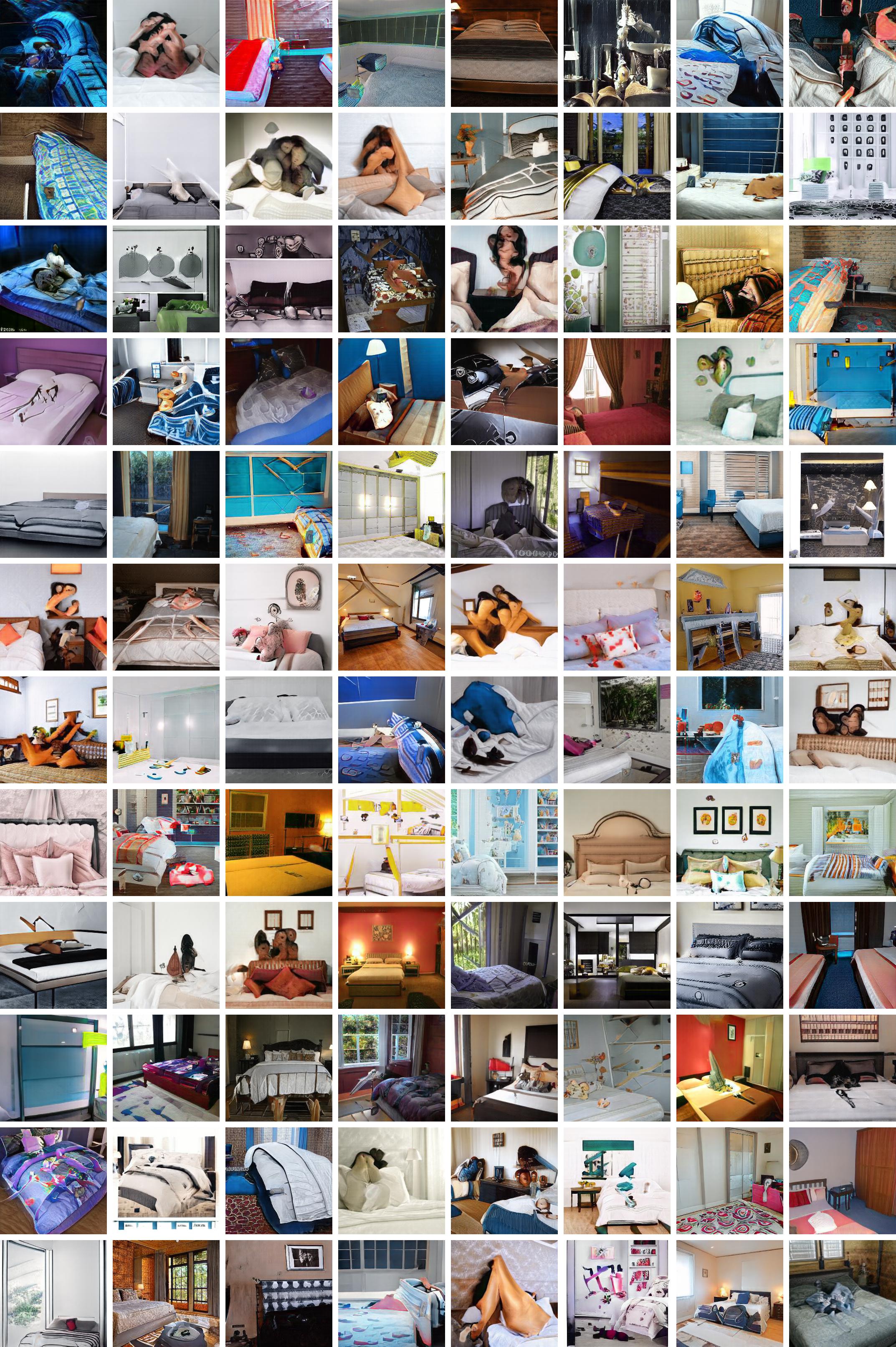}
 \vspace{-0.7cm}
\caption{The top-$96$ low quality generated images from StyleGAN trained on LSUN-bedroom.}
\label{fig:bedroom_worst}
\vspace{-0.2cm}
\end{figure*}

To demonstrate the superiority of our proposed approaches, we show more generated image quality assessment results. First, we apply the official StyleGAN~\cite{karras2019style} pretrained model to generate $5000$ images for each of the LSUN-cat, LSUN-car, and LSUN-bedroom dataset. Then we adopt our GMM-GIQA method to rank these generated images, and show the top-$96$ high quality(1.92\%) generated images in Figure~\ref{fig:cat}, Figure~\ref{fig:car}, Figure~\ref{fig:bedroom} for LSUN-cat, LSUN-car, and LSUN-bedroom, and the top-$96$ low quality(1.92\%) generated images in Figure~\ref{fig:cat_worst}, Figure~\ref{fig:car_worst}, Figure~\ref{fig:bedroom_worst}, respectively.

\clearpage

\end{document}